\documentclass{article}

\usepackage{microtype}
\usepackage{graphicx}
\usepackage{amsmath}  

\usepackage{subfigure}
\usepackage{booktabs} 

\usepackage{hyperref}

\usepackage[accepted]{mlsys2024}

\mlsystitlerunning{Learned Graph Rewriting with Equality Saturation: A New Paradigm in Relational Query Rewrite and Beyond}

\begin{document}

\twocolumn[
\mlsystitle{Learned Graph Rewriting with Equality Saturation: A New
Paradigm in Relational Query Rewrite and Beyond}

\begin{mlsysauthorlist}
\mlsysauthor{George-Octavian Bărbulescu}{cam}
\mlsysauthor{Taiyi Wang}{cam}
\mlsysauthor{Zak Singh}{cam}
\mlsysauthor{Eiko Yoneki}{cam}
\end{mlsysauthorlist}

\mlsysaffiliation{cam}{Department of Computer Science and Technology, University of Cambridge, United Kingdom}

\mlsyscorrespondingauthor{George-Octavian Bărbulescu}{gob24@cantab.ac.uk}

\mlsyskeywords{Query Rewrite, Graph Reinforcement Learning, Equality Saturation}

\vskip 0.3in

\begin{abstract}

Query rewrite systems perform graph substitutions using rewrite rules to generate optimal SQL query plans. Rewriting logical and physical relational query plans is proven to be an NP-hard sequential decision-making problem with a search space exponential in the number of rewrite rules. In this paper, we address the query rewrite problem by interleaving Equality Saturation and Graph Reinforcement Learning (RL). The proposed system, \textbf{Aurora}, rewrites relational queries by guiding Equality Saturation, a method from compiler literature to perform non-destructive graph rewriting, with a novel RL agent that embeds both the spatial structure of the query graph as well as the temporal dimension associated with the sequential construction of query plans. Our results show Graph Reinforcement Learning for non-destructive graph rewriting yields SQL plans orders of magnitude faster than existing equality saturation solvers, while also achieving competitive results against mainstream query optimisers.
\end{abstract}
]

\printAffiliationsAndNotice{} 

\section{Introduction}
\label{introduction}

Databases (DBs) are intricate systems for storing and managing data. To manipulate the data, DBs allow users to submit queries, 
 declarative statements through which the user specifies \textit{what} data to fetch or manipulate. The DB system manages \textit{how} the query is executed through query plans. Rewriting query plans is a fundamental problem in DB literature and a core driver for performance \cite{finance1991rule}. DBs employ query rewrite algorithms and rewrite rules (e.g. relational algebra rules) to transform the original database transaction into an equivalent and more efficient one. If executed effectively, the rewrite process can speed up the original transaction by orders of magnitude \cite{zhou2021learned}. The rewrite system takes as input a database query and projects it into a logical plan. Subsequently, the plan is rewired based on pre-defined rules (e.g. remove redundant operators) with two overarching goals in mind: 1) to produce a plan with faster execution time than the original, and 2) to ensure the output of the rewritten plan is equivalent to the original. 

Traditional query optimisers apply the rewrite rules in a predefined order, with each rewrite operating on the output of its predecessor \cite{begoli2018apache}. For example, PostgreSQL \cite{postgresql1996postgresql} rewrites queries in a top-down fashion. Graph substitutions are performed sequentially starting from the query plan root node, based on pattern matching. The inherent limitation of this approach is the potential to fall into a local optimum when the rewrite rules have inter-dependencies. Finding the optimal order to apply query rewrite rules is generally NP-hard \cite{finance1991rule, pirahesh1992extensible}. An alternative approach to mainstream query optimisers is to apply all rules (i.e. graph substitutions) at once and subsequently pick the best query plan from the conglomerate. This paradigm is known as equality saturation, and it has been successfully applied in a variety of optimisation problems \cite{wang2020spores, yang2021equality}. However, a naive application of equality saturation to rewrite SQL queries in the general case is set for failure. While equality saturation as proposed by \cite{tate2009equality} employs equality graphs (e-graphs) to store a potentially infinite number of expressions (e.g. query plans) within finite memory, the algorithm's computational requirements remain impractical for local and cloud DB deployments with limited resources. The limitations of equality saturation for query rewrite are not solely tied to infeasible memory requirements but also hamper the algorithm's execution time which is reflected in the end-to-end latency of SQL queries. 

These limitations underscore the need for an alternative approach capable of constraining the non-destructive graph rewriting algorithm within reasonable memory and time limits. On their own, e-graphs help solve one of the pressing challenges in the query rewrite literature, which looks at how to represent a large number of query plans \cite{zhou2021learned}. Through \textbf{Aurora}, we aim to solve the remaining challenge of identifying optimal query plans given a set of rewrite rules in relational DBs. We introduce a new Graph Reinforcement Learning agent that is able to navigate the complex search space of e-graphs by using a mix of Graph Neural Networks (GNNs) and Recurrent Neural Networks (RNNs). We show that by guiding the construction of query e-graphs with RL, \textbf{Aurora} outperforms existing state-of-the-art equality saturation solvers, returning order-of-magnitude faster query plans within a service-level agreement (SLA) on memory requirements. Furthermore, the RL-based equality saturation paradigm we use to rewrite relational query plans can be adapted to other programming languages than SQL with microscopic changes. 

Concretely, our contributions are:
\begin{enumerate}
    \item A pioneering approach to the query rewrite problem using Deep Reinforcement Learning (DRL) within the context of equality saturation.
    \item The design and implementation of Aurora, a database query optimizer that leverages RL-based equality saturation. The methodology encompasses: a) a novel spatio-temporal DRL algorithm for solving combinatorial problems over graphs; and b) an integer linear programming (ILP) formula for extracting query plans from e-graphs.
    \item An extensive experimental evaluation of Aurora against the state-of-the-art equality saturation solvers. We show that Aurora reduces the cost of query plans by orders of magnitude compared to the baselines, and it's competitive against traditional DB optimisers.
\end{enumerate}

\section{Background}

This section introduces equality graphs, as a method to represent a conglomerate of equivalent expressions, the equality saturation algorithm as initially proposed by \cite{tate2009equality}, and reinforcement learning (RL) as a method to optimise sequential decision-making.  

\subsection{Equality Graphs}

Equality graphs (e-graphs) are data structures that efficiently represent congruence relations \cite{nelson1980fast, nieuwenhuis2005proof}. E-graphs, initially developed for automated theorem provers (ATPs) \cite{willsey2021egg}, extend the union-find method proposed by \cite{tarjan1975efficiency} to efficiently encode expressions in equivalence classes. A core property of the e-graph is the ability to close the equivalence relation under congruence (e.g. if $a \equiv b  \rightarrow f(a) \equiv f(b)$). 

E-graphs have become ubiquitous in rewrite-driven compiler optimisations, as part of a technique coined equality saturation \cite{tate2009equality, wang2020spores, wang2022optimizing, yang2021equality}. Equality saturation receives an input expression (e.g. $a+b$) and stores it in an e-graph. Subsequently, the e-graph is expanded iteratively by applying rewrite rules over the information it holds. An important remark is that growing the e-graph is a purely additive process - the rewrites solely add information instead of changing the original input expression or the expressions that follow from expanding the e-graph. Upon reaching \textit{saturation}, the e-graph is deemed to represent all possible equivalent expressions that can be derived from the rewrite rules and the input expression. 

\subsection{Equality Saturation}

\begin{figure}[t]
  \centering
  \includegraphics[width=\linewidth]{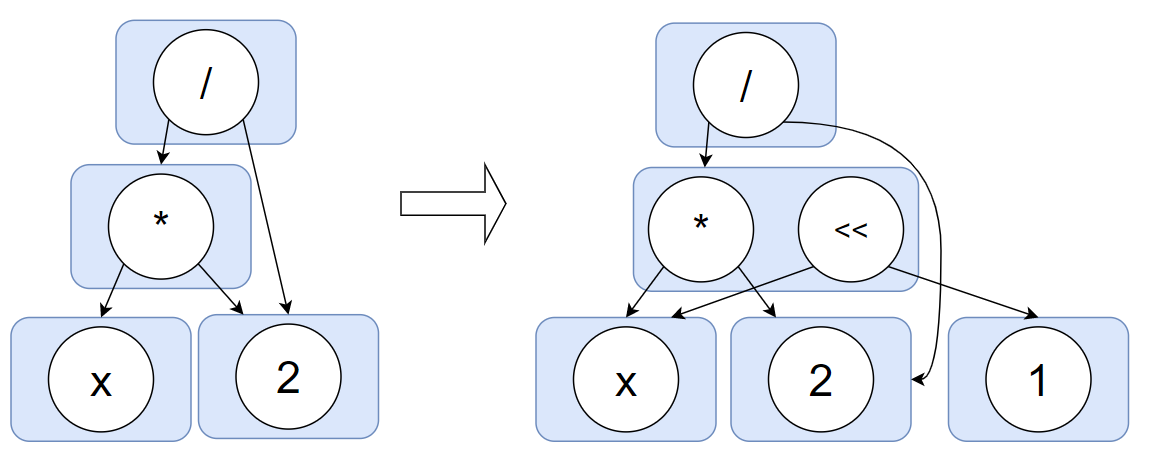}
  \caption{E-graph construction. The e-graph undergoes rewrite $x*2 \rightarrow x<<1$. Figure adapted from \cite{willsey2021egg}.}
  \label{fig:math_example}
\end{figure}

Figure \ref{fig:math_example} shows the construction of an e-graph for mathematical optimisation. The rewrite rule $x*2 \rightarrow x<<1$ only adds information to the original expression $(x*2)/2$. If there exists a rule such that $(x*2)/2 \rightarrow x$, this rule can be subsequently fired over the e-graph to produce an optimal plan in terms of the number of operations. In a classic term rewrite system, the left-hand side expression $(x*2)/2$ would have been obfuscated by $(x<<1)/2$, thus leading to an ordering problem. An e-graph consists of e-classes (blue boxes) and e-nodes (circles). An e-class contains one or more e-nodes. Picking any e-node from an e-class is ensured to produce a correct and equivalent expression in reference to the original. 

In equality saturation, all available rewrite rules are applied sequentially until the e-graph no longer changes. Once saturation is reached, the optimal form of an expression can be extracted from the e-graph via a greedy search or a more sophisticated algorithm such as integer linear programming (ILP) \cite{yang2021equality}. In Section \ref{sec:challenge_es}, we discuss why equality saturation cannot be trivially applied to complex programming languages and problems such as query rewriting.  

\subsection{Reinforcement Learning} \label{sec:rl}
Reinforcement Learning (RL)  encompasses a class of techniques for optimising sequential decision-making processes. One formalisation of RL is Markov Decision Processes (MDPs) \cite{sutton2018reinforcement}, which are instruments for characterising decision-making problems over a discrete time dimension \cite{puterman2014markov}. The decisions are taken by an agent that acts over a Markov chain \cite{norris1998markov}, an environment where the agent's future state depends only on its previous state and the action taken \cite{frydenberg1990chain}.  

Formally, an MDP is a 4-tuple $(S, A, T, R)$ which describes the state space $S$, the action space $A$, the dynamics of the environment $T$, and a reward signal $R$, respectively. To learn a \textit{policy} for the agent to follow, Artificial Neural Networks (ANNs) are typically used to generalise across large state-action (S-A) spaces, which is known as Deep Reinforcement Learning (DRL).

Aurora aims to learn a policy over the space of e-graphs (S) constructed from input query plans, where the action space A is defined by the set of available rewrite rules (e.g. relational algebra rules) and the reward is the expressiveness of the e-graph (the latency of the extracted query plan) derived within an SLA constraint such as the memory required to store the e-graph. This formalisation serves as the core intuition behind Aurora.

\section{Problem Overview} 

\subsection{Challenges in Query Rewrite} \label{sec:challenge_qr}

\begin{figure}[t]
  \centering
  \includegraphics[scale=0.2]{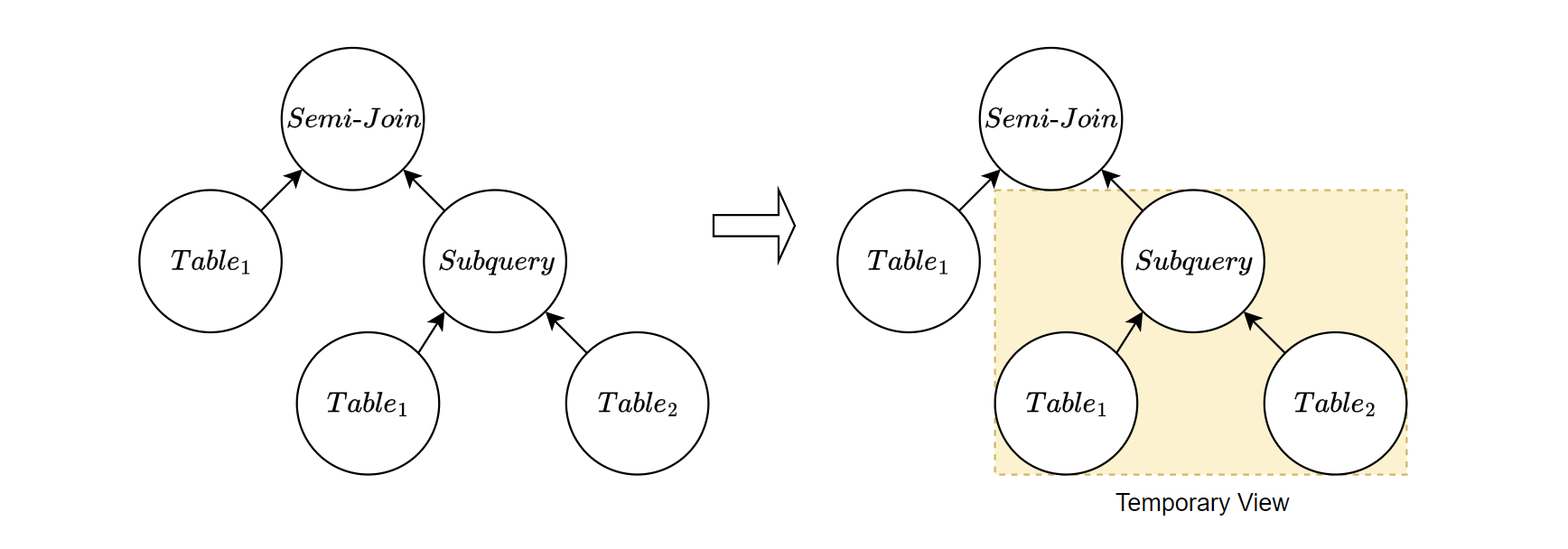}
  \caption{The original SQL plan (left) is optimised in a top-down fashion by creating a temporary view over the subquery (right).}
  \label{fig:example_qc}
\end{figure}

Query rewriting is an NP-hard problem \cite{pirahesh1992extensible, finance1991rule} due to the exponential search space. Database engines project SQL queries into logical plans, which are subsequently optimised based on hand-crafted rules (e.g. perform filtering foremost in the plan). The application of rules (i.e. substitutions) commonly follows an ordering scheme (e.g. arbitrary, heuristic) \cite{zhou2021learned}. In turn, this can lead to suboptimal query plans, as rewrite rules may have inter-dependencies. Concretely, substituting parts of the original plan may render more prosperous future rewrites inapplicable. 

 To build our intuition, we refer to a study case provided by the authors of \cite{zhou2021learned}, who investigate the inner workings of PostgreSQL. PostgreSQL facilitates database transactions in a top-down manner. The engine traverses the logical query plan starting from the root and performs local rewrites iteratively. Figure \ref{fig:example_qc} renders an example SQL query for which a top-down optimiser finds a sub-optimal plan. The original plan tree is representative of SQL templates such as \textit{``select \_ from $table_1$ with condition over (select \_ from $table_1$, $table_2$ where \_)"}. Once the optimiser reaches the $Subquery$ node in the plan, it naively creates an inline view. Consequently, further optimisations are disabled over the view. By disregarding the characteristics of the leaf nodes, a top-down application of rewrite rules will not take advantage of the correlation between the subquery and the outer query (note the same relation - $Table_1$ - is scanned twice). Ideally, the subquery is pulled up as a join to prevent a redundant scan operation. In practice, this simple optimisation reduces the execution time by a factor of $600$ as per the original authors \cite{zhou2021learned}. 

\subsection{Challenges in Applying Equality Saturation} \label{sec:challenge_es}

Equality Saturation (ES) starts from the promise to address the ordering problem in classic rewrite-based systems. Different from term rewriting systems, ES employs e-graphs to perform non-destructive graph rewriting. We investigate the theoretical and practical limitations of equality saturation. 

One of the concerns raised in literature is the prominence of the algorithm to run into \textit{``explosions"}. We coin an explosion a soar in the number of nodes in the e-graph in the aftermath of a suboptimal rewrite rule application. As a concrete example, \cite{yang2021equality} investigate this dimension in the context of tensor graph optimisation, showing that rewrite rules can quickly grow the e-graph to the upper limit of $50k$ (thousands) nodes they set. The consequences of a large e-graph are twofold: 1) the data structure may not fit into memory in complex domains where explosions are frequent, and 2) it hampers the time to extract the optimal expression from the e-graph. The latter point is a pressing concern in systems that provide real-time feedback, such as database query engines. In this scenario, the overhead associated with growing the e-graph and extracting the optimal expression is included in the end-to-end latency of a query execution. In this case, constraining equality saturation with RL becomes mandatory to achieve competitive latency. 

Finally, Willsey et al. \cite{willsey2021egg} render a scenario in which equality saturation is characterised by non-termination, where a purely \textit{destructive} approach would terminate. In this context, we denote destructive any rewrite system which performs substitutions. The condition of non-termination is linked to an unbounded e-graph expansion, one that never saturates. Identifying an instance of this for SQL or relational algebra remains a question for future research, but one that's tackled inherently by Aurora. 

\begin{figure*}[t]
  \centering
  \includegraphics[scale=0.3]{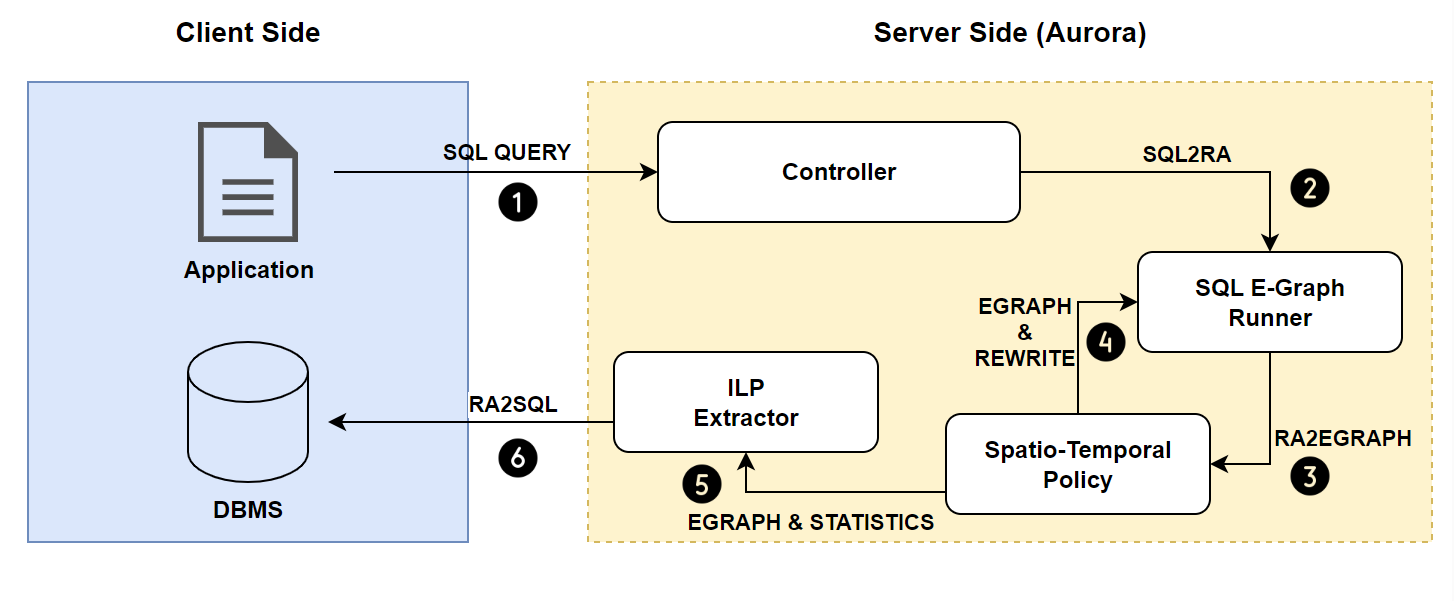}
  \caption{A 6-step schematic workflow of Aurora}
  \label{fig:workflow}
\end{figure*}

\section{Solution Overview}

Aurora aims to jointly address the challenges underpinning query rewrite and the limitations associated with existing equality saturation solvers. The usage of e-graphs is motivated by the large number of equivalent plans that can be derived from relational rules and beyond. To perform query rewrite with e-graphs within reasonable planning latency and memory bounds, we introduce RL into the equality saturation paradigm. We show this leads to query plans order of magnitude faster than the ones derived by existing ES solvers. Figure \ref{fig:workflow} renders Aurora's workflow. The system's actions are clustered into five overarching steps, with an optional sixth step that shows how the framework can be used as a SaaS for mainstream databases (e.g. PostgreSQL).

\begin{itemize}

    \item \textit{Step 1:} Aurora's \textit{controller} receives a SQL query. 
    \item \textit{Step 2:} The SQL query is parsed into an \textit{extended} relational algebra (RA) expression, with the caveat that the logical plan can include physical operators (e.g. hash-join, merge-join). The motivation behind the design is to jointly rewrite the query and to set flags for physical execution.
    \item \textit{Step 3:} The derived RA expression is used as input for a new e-graph. The expression is split into e-classes and e-node operators (e.g. project, filter). 
    \item \textit{Step 4:} The RL agent guides the expansion of the e-graph. To achieve this, we apply pattern-based rewrite rules (e.g. $join(a,b) \rightarrow join(b,a)$) to generate new e-nodes and e-classes. The logic is encapsulated in an E-Graph Runner which uses e-graphs good (egg) \cite{willsey2021egg}, the state-of-the-art equality saturation implementation. 
    \item \textit{Step 5:} The e-graph and statistics from the database catalogue (e.g. cardinality, selectivity) are sent to 
    an \textit{extractor} module. The module extracts the best query from the e-graph using integer linear programming (ILP) and a cost heuristic we explain in Section \ref{sec:extraction}.
    \item \textit{Step 6:} The extracted relational plan can be converted back into a SQL dialect, for instance, to be used in mainstream databases such as PostgreSQL. This final step is optional in Aurora. The reason is that the framework acts as an end-to-end SQL optimiser with access to direct execution in Risinglight \cite{rising}, a Rust-based DBMS for analytical queries. 
\end{itemize}

\section{Reinforcement Learning for Equality Saturation} \label{sec:spatio-temporal}

Having established existing equality saturation solvers cannot be trivially used to generate good query plans due to challenges such as memory requirements and latency considerations (see Section \ref{sec:exp_mot} for experimental results), we introduce RL as a method to execute equality saturation efficiently. Intuitively, the goal is to encode in the e-graph as many competitive equivalent query plans to the original as possible within a limited time and memory budget. 

\subsection{Equality Saturation as Markov Decision Process} 
\label{sec:mdp_es}

This section formalises equality saturation for query rewrite through the lens of a Markov Decision Process (MDP). 

\subsubsection{State}

\begin{figure}[t]
  \centering
  \includegraphics[width=\linewidth]{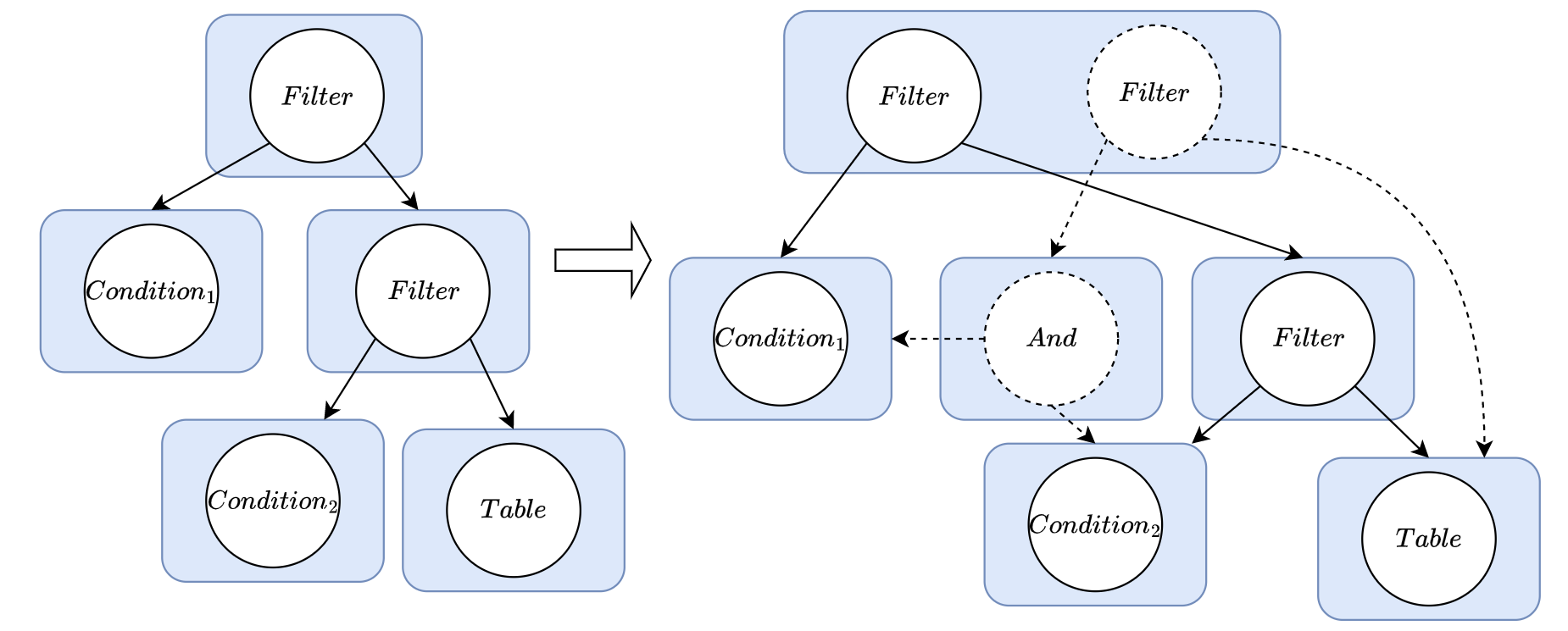}
  \caption{Example e-graph for RA expression $\pi(\sigma_{c1}(\sigma_{c2}(t)))$. By using the rewrite rule ``$\sigma (c_1, \sigma (c_2, t) )\rightarrow \sigma( c_1 \land c_2, t)$", the query e-graph adds the equivalent plan $\pi(\sigma_{c1 \land c2}(t)))$. The circles represent e-node operators and the blue boxes represent e-classes}
  \label{fig:equivalence}
\end{figure}

\textbf{Example:} The SQL query \textit{``select * from (select * from t where c2=True) where c1=True"} is projected into relational algebra and rewritten by the query planner as $\pi(\sigma_{c1}(\sigma_{c2}(t)))$. An example is provided in Figure \ref{fig:equivalence}. 

\begin{figure}[t]
  \centering
  \includegraphics[width=\linewidth]{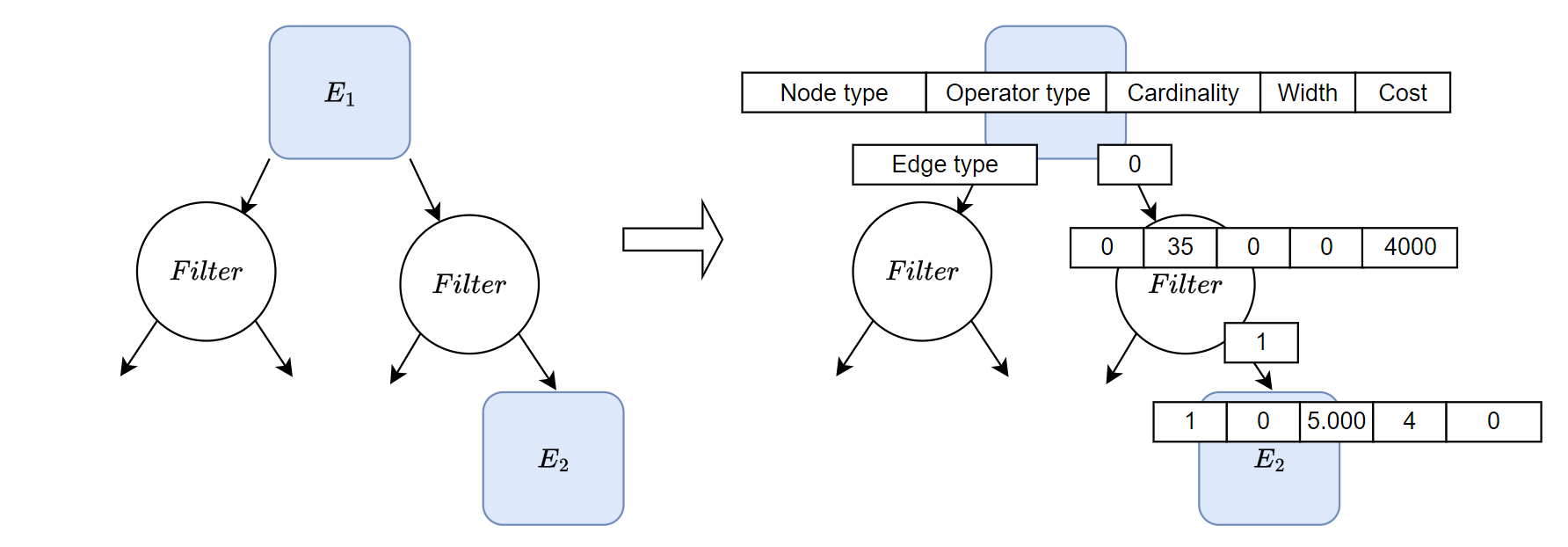}
  \caption{Example e-graph encoding. The encoding scheme covers both the nodes as well as the edges in the e-graph}
  \label{fig:encoding}
\end{figure}

We take a slice from the rewritten e-graph rendered in Figure \ref{fig:equivalence}, which shows the top-most equivalence class (e-class $E1$), and two `filter' operators (Figure \ref{fig:encoding}).

Our proposed node encoding over the query e-graph is a 5-tuple (Table \ref{tab:encoding_q}). The \textit{node type} distinguishes between e-nodes and e-classes. An e-class is a pointer to a set of e-nodes (e.g. filter, join). The \textit{operator type} is the core component of the encoding scheme. We note the rewrite actions over the e-graph are solely facilitated by the presence of certain operator patterns (e.g. a filter operation before a join operation). We propose encoding statistics from the target database management system. If the node is an equivalence class, then the feature vector stores the \textit{cardinality} (number of rows) and the \textit{width} (the number of columns/data bytes of a tuple) returned by its e-nodes. It is important to make a distinction here between e-classes and e-nodes. The e-nodes are executable operators (e.g. filter) and they can only be described through operator costs. To estimate the cost of an operator, we merely need to know the size of the output of each of its children. We remark that e-classes can be described through cardinality and width, which are oblivious of the e-node picked.  Finally, we add edge attributes to distinguish between e-node to e-class edges and vice versa. 

\begin{table}[t]
  \caption{The node encoding of the query equality graph} 
  \label{tab:encoding_q}
  \begin{tabular}{lp{0.7\linewidth}} 
    \toprule
    Feature & Details\\
    \midrule
    node type& One-hot encoding over node types in e-graphs. A node in the e-graph can either be an equivalence class or an e-node.  \\ 
    operator type& One-hot encoding over the operators. The number of operators spanned by the extended RA language is 64. \\ 
    cardinality& The number of records in the e-class\\
    width& The number of columns in the e-class\\
    cost& The estimated execution cost\\
    \bottomrule
  \end{tabular}
\end{table}

\subsubsection{Action \& Transition Function}

The action space is the set of query plan rewrite rules. Aurora uses 37 rules that encompass RA, mathematical, and boolean rewrite rules, by expanding Risinglight \cite{rising}. 

Aligned with previous work \cite{omelette}, we use a \textit{Reset} action. The \textit{Reset} action extracts the optimal plan from the current e-graph and wraps it into a new e-graph. The aim is two-fold: 1) to induce a heuristic-like effect during e-graph construction, and 2) to reduce the size of the e-graph while maintaining opportunities for future optimisation. 

\subsubsection{Reward Signal}

The reward system encourages latency improvements by benchmarking the final query plan extracted from an e-graph in an optimisation episode. Positive rewards are computed by measuring the cost improvement between the previous and newly-extracted plans from the e-graph. In terms of negative rewards, the agent is penalised when taking actions that are either saturated (i.e., don't add new nodes) or lead to large memory overhead.

\subsection{Spatio-temporal Reinforcement Learning}

\begin{figure}[t]
  \centering
  \includegraphics[scale=0.27]{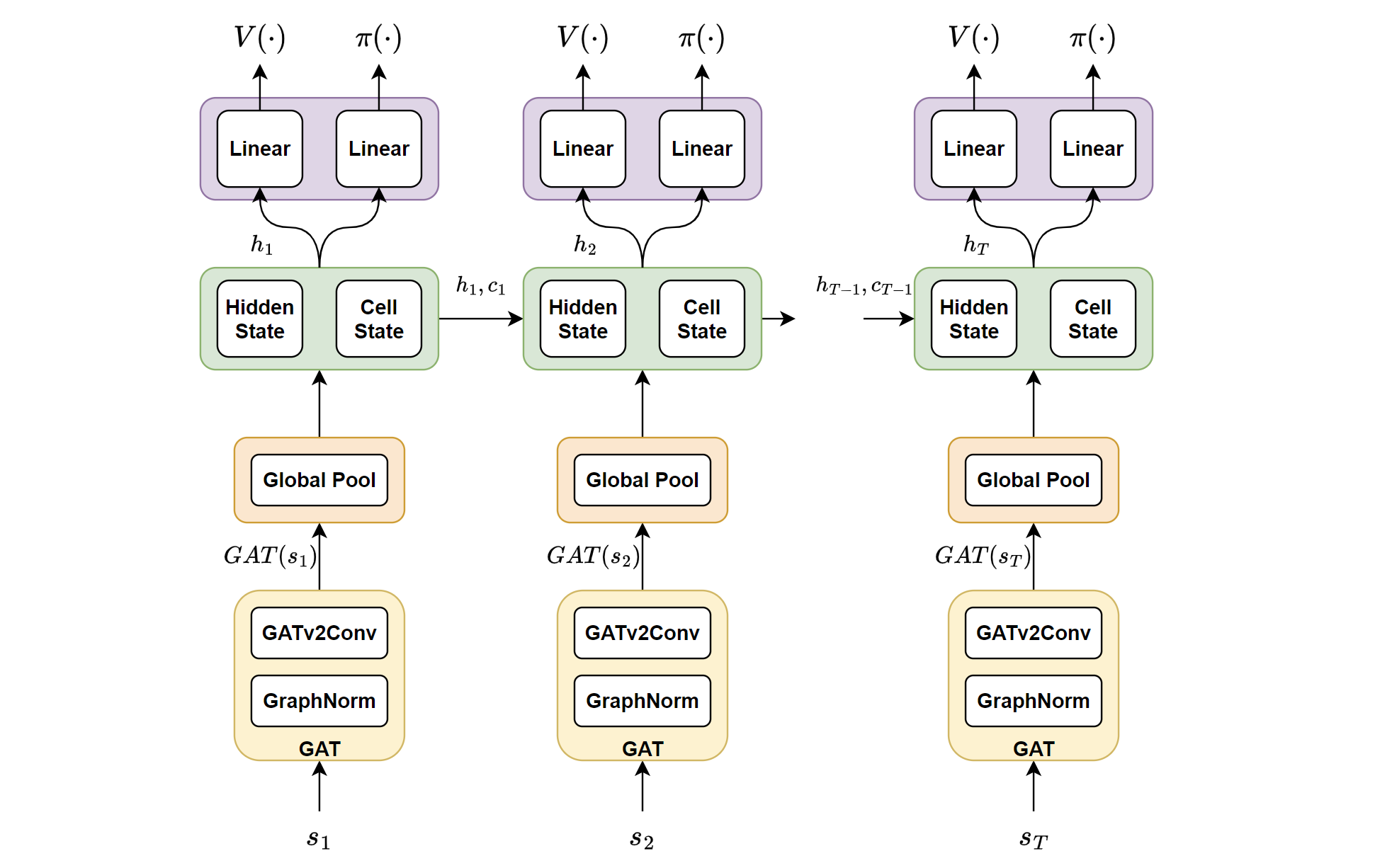}
  \caption{The architecture of the spatio-temporal RL agent. It features a Graph Attention Network layer, a global add pool, a Long Short-Term Memory layer and two linear heads for actor and critic, respectively.}
  \label{fig:drl-fig}
\end{figure}

We propose an actor-critic RL agent \cite{sutton2018reinforcement} to learn the environment (Figure \ref{fig:drl-fig}). We use knowledge from Graph Neural Networks (GNNs) \cite{zhou2020graph} and Recurrent Neural Networks (RNNs) \cite{medsker2001recurrent} to learn the dynamics of expanding query e-graphs. 

\subsubsection{Learning the Spatial Representation}
In equality saturation, the e-graph serves two interleaved purposes: 1) it efficiently represents multiple versions of compiler plans, and 2) it guides the rewrite process. The layout of the e-graph guides which rewrite rules can be applied over its structure based on pattern matching. Given some rewrite rules are more beneficial than others, it is paramount to learn node neighbourhood patterns, warranting an attention-based method.

GNNs are learning methods for graph-structured data \cite{zhou2020graph}. The key idea behind GNNs is to use pairwise message passing between graph nodes to update their latent representations. The Graph Attention Network (GAT), introduced by Velickovic et al. \cite{velickovic2017graph}, is an attention-based architecture for graph-structured data. A key advantage of GATs over other types of graph learning methods, such as Graph Convolutional Networks (GCN) \cite{kipf2016semi}, is their ability to automatically infer the importance of graph substructures. In our proposed agent architecture, a 3-layer GAT is introduced to model a spatial representation of the encoded e-graph. 

\subsubsection{Learning the Temporal Link}
Recurrent neural networks (RNN) \cite{medsker2001recurrent} are neural networks for learning sequential data. The key innovation behind RNNs is the ability to memorize previous steps to influence subsequent inference. We leverage an RNN to model the trajectory in the e-graph construction MDP. The motivation is two-fold. First, the RNN is employed to capture the purely additive nature of rewriting e-graphs (e-nodes and some e-classes will not change). Second, we aim to mitigate some of the expressiveness pitfalls of the graph encoder. As an example, a global mean pool function may project two distinct e-graphs into the same latent representation, leading to a sub-optimal policy.  

Aurora uses a Long Short-Term Memory (LSTM) \cite{sherstinsky2020fundamentals, hochreiter1997long} to prevent the vanishing gradient problem \cite{hochreiter1998vanishing} over long equality saturation episodes. The LSTM is formally defined below:
\begin{equation} \label{acum_eq}
    c_t = f_t \odot c_{t-1} + i_t \odot g_t,\quad h_t = o_t \odot \tanh(c_t)
\end{equation}

In Equation \ref{acum_eq}, $c_t$ collects information from the state space. There are two gates: an input gate denoted $i_t$ and a forget gate denoted $f_t$. If the input gate is on, then the newly learnt representation $g_t$ over the current state is stored. If the forget gate is on, then some information from the preceding cell $c_{t-1}$ is discarded. Finally, the information from the cell is passed to the hidden state $h_t$, with a dependency on the output gate $o_t$.

\subsubsection{Spatio-temporal Proximal Policy Optimisation}

Aurora extends PPO \cite{schulman2017proximal}, a policy gradient RL method, to perform gradient updates on sequential training data. Additionally, we use the best practices from previous literature on RNNs for RL \cite{huang202237, piao2019energy} and instantiate the LSTM hidden and memory cells with values of zero, which are reset at the beginning of each RL episode. The \textit{actor} and the \textit{critic} networks ($V$ and $\pi$ in Figure \ref{fig:drl-fig}) produce a probability distribution of rewrite actions and the value of an e-graph state encoding, respectively. In a closed form, these can be computed as follows: 

\begin{equation}
    h_t, c_t = lstm(pool(gat(s_t))), h_{t-1}, c_{t-1})
\end{equation}
\begin{equation} \label{eq_vf}
    \pi_{\theta_k} = f_{actor}(h_t), \quad V_{\theta_k} = f_{critic}(h_t)
\end{equation}

The computations above, in which $\theta_k$ represents the parameters of the entire deep model after $k$ training iterations, and $s_t$ the e-graph encoding, respectively, can be used during inference to solve a new equality saturation instance.

\section{Query Plan Extraction} 
\label{sec:extraction}

The final instrument in Aurora is the e-graph extraction routine (step 5 in Figure \ref{fig:workflow}), assigned to extract the optimal plan from the query e-graph. It is important to remember the e-graph is an efficient encoding of multiple equivalent query plans.

Extracting the optimal plan from the e-graph is non-trivial in the database scenario. While previous work from the realm mainly focused on the size of the expression (i.e., the number of nodes in the query plan) as a cost function \cite{omelette}, we deem this provides no intuition regarding the performance of a query plan. To this end, it is paramount to devise a cost function over the query e-graph that can be linked back to the end-to-end latency of a query plan. In our implementation, we try variations of the cost model proposed by \cite{rising}. 

First, a proxy for the size of the data that flows through the query plan is defined. Define $n$ the number of tuples and $m$ the number of columns in a target table. The size of the data is proportional to the intermediate result returned by an operator. For example, the intermediary output returned by a \textit{scan} operation over the table has a size of $n \cdot m$. If the \textit{scan} operator is followed by a \textit{filter} operation, which without loss of generality we assume reduces the number of tuples by $50\%$, then the new intermediary result will have a size  $\frac{n}{2} \cdot m$. 

As shown in Section \ref{sec:mdp_es}, the size of the intermediary results can be stored in e-classes at runtime. This is because all operators (e-nodes) from an e-class should produce equivalent outputs. 

Second, an operator cost function is implemented over the size of the intermediary result. Each e-node is either a plan operator, a column, or a table. We use the asymptotic costs from DB literature to assign costs. For instance, if a table has $n$ tuples and $m$ columns, then a \textit{scan} operation is assumed to cost $\mathcal{O}(nm)$. Finally, the plan is extracted from the e-graph using integer linear programming to minimise the cost heuristic above, a popular technique in equality saturation \cite{yang2021equality}.

\section{Evaluation} \label{ch:eval}
This section describes our experimental evaluation of Aurora. Its performance as an end-to-end query rewrite mechanism is evaluated on a complex suite of analytical SQL queries. The results are evaluated against e-graphs good \cite{willsey2021egg}, the state-of-the-art equality saturation solver, Omelette \cite{omelette}, an RL-based equality saturation solver, and a pseudo-random agent. Each solver has access to the same set of rewrite rules Aurora uses. 

\subsection{Experimental Setup}

\subsubsection{Testbeds}
There are two database management systems (DBMSs) used in our experimental analysis. Evaluating Aurora as a competitive equality saturation solver is done by rewriting queries in Risinglight against the equality saturation baselines. To further validate Aurora's utility as a query rewriting tool, we also benchmark it against PostgreSQL's in-built optimizer, reinforcing Aurora's value as a robust query optimiser. 

\begin{enumerate}
    \item \textbf{Risinglight} \cite{rising} is a DBMS implemented in Rust. It exhibits a database query engine that independently confines rewrite rules. Aurora expands the set of rewrite rules from Risinglight and extends the engine to support user-defined rewrite orders. The plans devised by Aurora can be directly executed against Risinglight's storage system without converting the plan back into a SQL dialect. The usage of Risinglight as a testbed is further motivated by the smooth integration with e-graphs good, which we show to be a robust alternative to mainstream query engines.  
    \item \textbf{PostgreSQL} \cite{postgresql1996postgresql} is a popular mainstream DBMS with wide applications in both industry as well as research. To execute queries on PostgreSQL, the query plans devised by Aurora are translated back into the Postgres dialect. 
\end{enumerate}

\subsubsection{Benchmarks \& Metrics}

Aurora has to be effective at rewriting complex database transactions, such as multi-join analytical queries. To achieve the desired level of rewrite difficulty, we experiment with the TPC-H database benchmark \cite{poess2000new}. The TPC-H benchmark is an online analytical processing (OLAP) workload. The database schema contains 62 columns across eight tables. The queries simulate decision support transactions associated with warehousing environments. For our experimental analysis, we subtract a subset of queries from the TPC-H suite that exhibits high rewrite potential and is reflective of the intricacies of the TPC-H benchmark.

We evaluate Aurora as a query rewrite system through query planning and execution latency. Furthermore, we evaluate the performance of our e-graph cost model (see Section \ref{sec:extraction}) by measuring the estimated cost against the real latency.

\subsubsection{Training \& Hardware}

For each combinatorial optimisation task, Aurora and Omelette are trained for $200$k steps, with a horizon of 200 steps. Unless otherwise noted, all the experimental results are averaged over $5$ seeds. The same seeds are used for all algorithms. For any given seed, given the trained policies are stochastic, the pre-trained agents are rolled out in the environment for $100$ roll-outs, collecting only the best results. The same applies to the heuristic agent. This provides reasonable statistical significance across the results. E-graphs good is run only once given it is deterministic.

The system used for training exhibits a Tesla P100 GPU with 16GB of VRAM, 24 logical CPUs, and 64GB of memory.

\subsection{Performance Results}
\begin{figure}[t]
\includegraphics[scale=0.4]{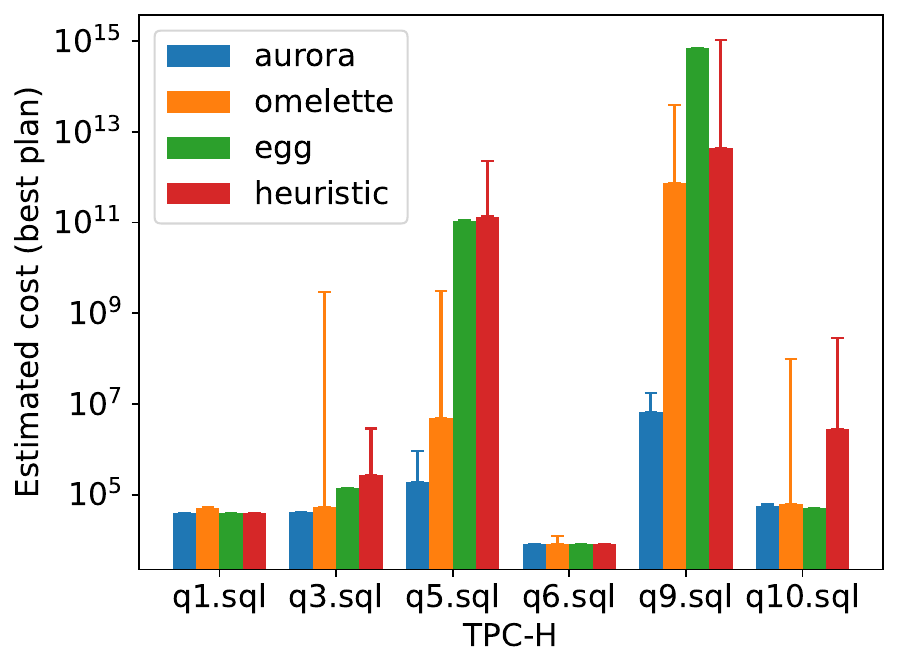} 
\centering
\caption{The median cost of the extracted plan by algorithm over 5 seeds. The whiskers represent the maximums. Lower cost is better.}
\label{fig:estimated_performance}
\end{figure}

\begin{figure}
\includegraphics[scale=0.40]{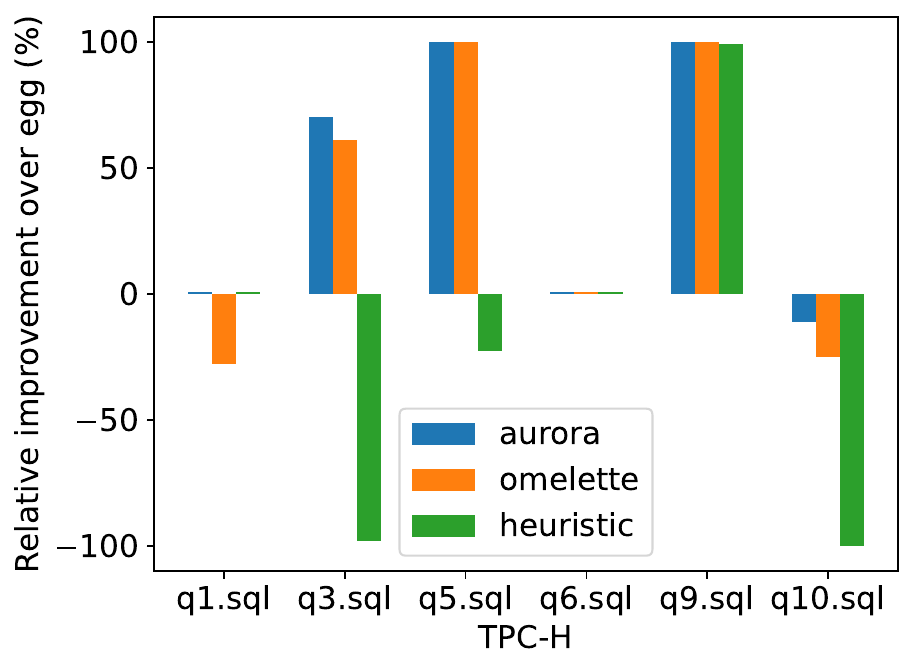} 
\centering
\caption{The relative improvement over \textbf{egg}, the state-of-the-art equality saturation solver. The negative outcome is bounded at -100\%. Higher relative improvement is better.}
\label{fig:relative_performance}
\end{figure}

\begin{figure*}[ht]
\includegraphics[scale=0.4]{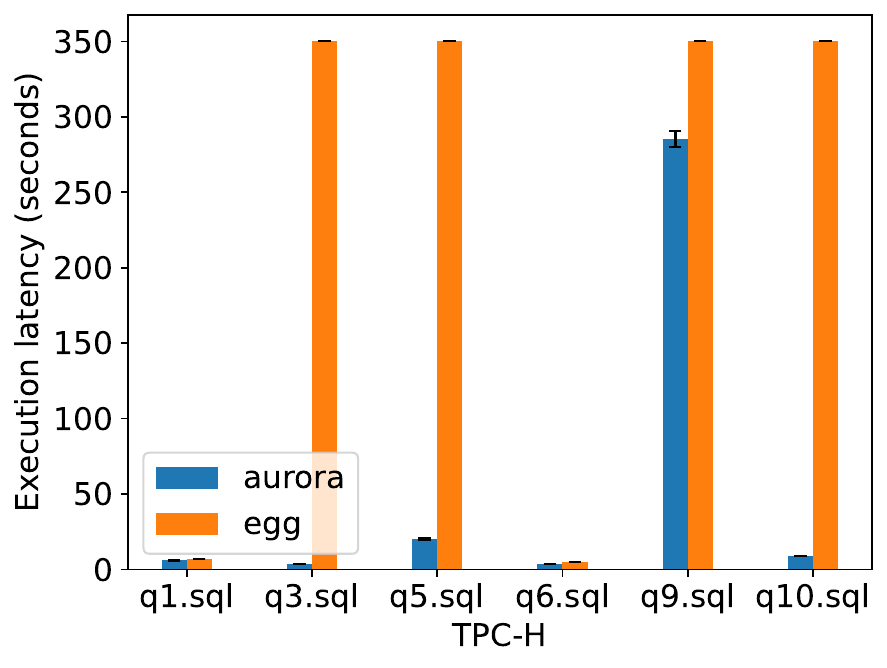} 
\includegraphics[scale=0.4]{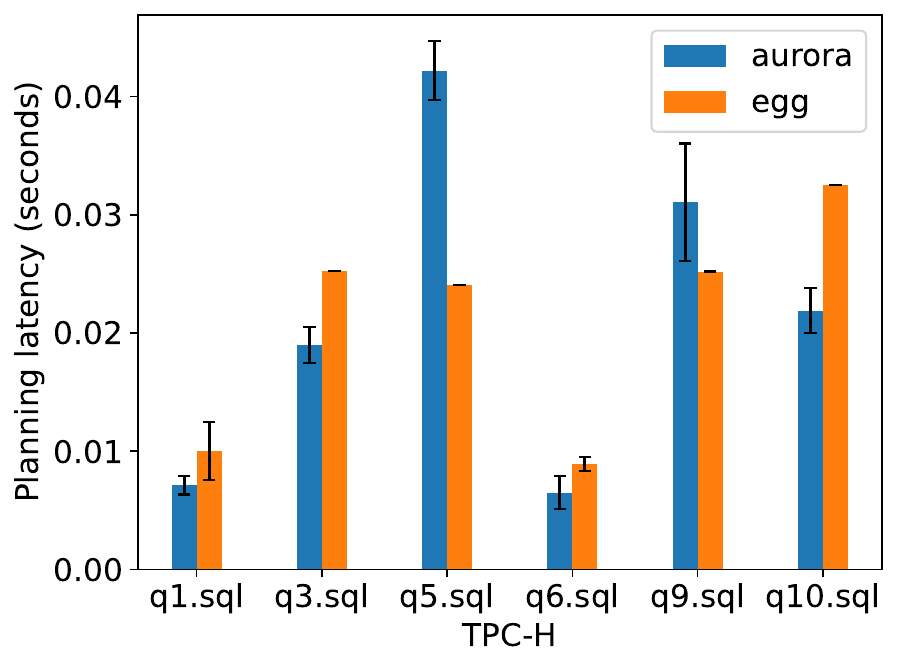} 
\centering
\caption{\textbf{Left:} the execution latency of the extracted SQL plan with a timeout set at 350 seconds. \textbf{Right:} the planning latency. All metrics are averaged over 5 trials. In both cases, lower is better.}
\label{fig:actual_performance}
\end{figure*}

We first prove Aurora is able to rewrite query plans exhibiting lower costs than existing equality saturation solvers. We use Risinglight for direct execution and set an upper bound on the number of nodes in the e-graph to $1000$ nodes to serve as the memory constraint in the optimisation problem. 

Figure \ref{fig:estimated_performance} renders the ability to rewrite queries across our baselines. The estimated cost is computed by extracting the final plan from the e-graph. It is evident from the plot that Aurora discovers query plans that are orders of magnitude lower in cost than the baselines. 

We observe that a random application of rules yields higher-cost plans than egg. The significance of this result is two-fold. First, it shows that some structure is needed to solve the query e-graph construction problem. The lack of structure in random search plays a detrimental role in solving the query rewrite combinatorial problem, by making it infeasible for the agent to \textit{guess} a competitive rule ordering. Second, it warrants that RL is a competitive choice for optimising query plans over the space spanned by e-graphs. This follows from the performance gap between the RL agents and heuristic search.

\begin{figure*}
\includegraphics[scale=0.32]{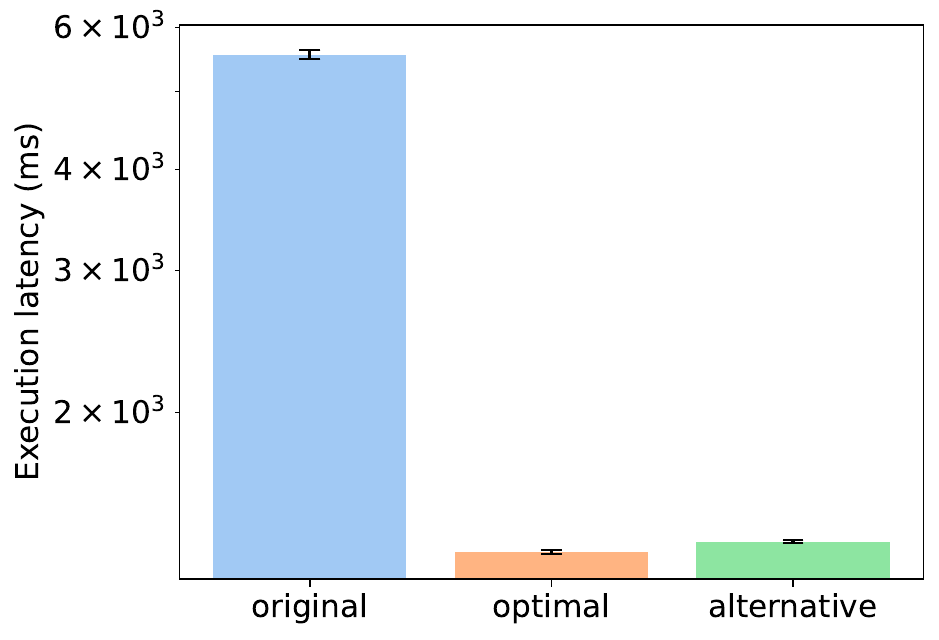} 
\includegraphics[scale=0.32]{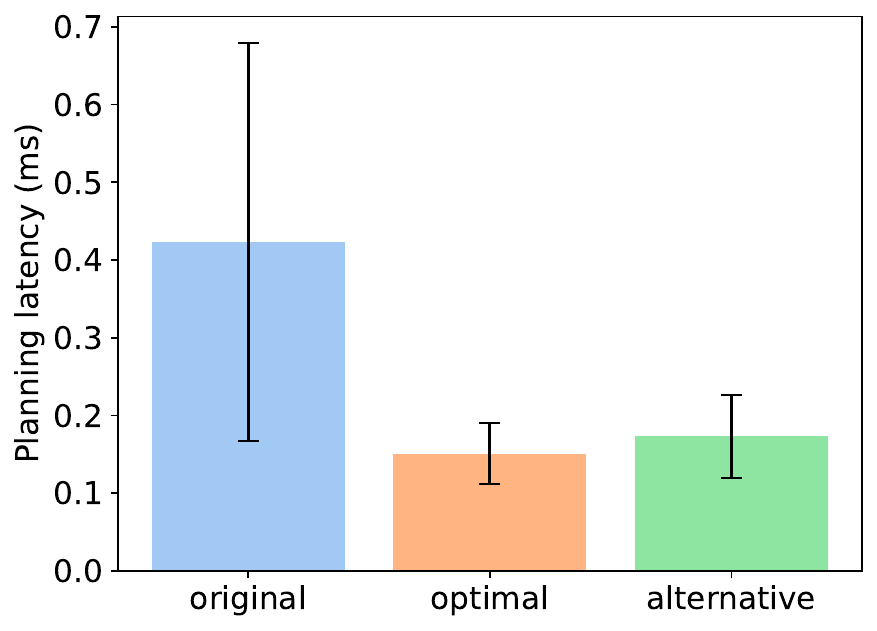} 
\includegraphics[scale=0.32]{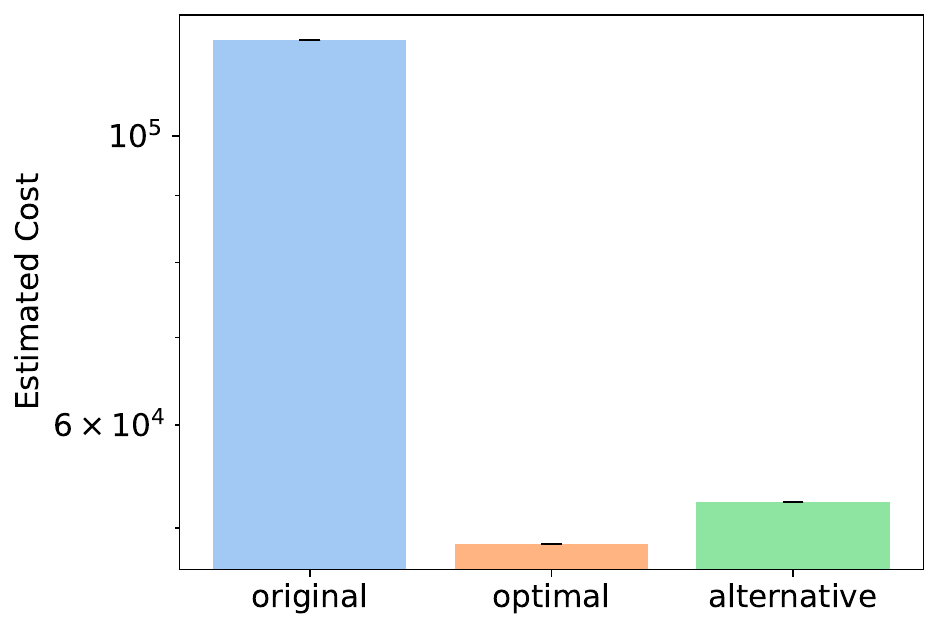} 
\centering
\caption{From left to right: the execution latency, the planning latency, and the estimated cost (PostgreSQL) of the queries from Table \ref{tab:pgsql_case} (5 trials). Lower is better in all cases.}
\label{fig:postgres}
\end{figure*}
Figure \ref{fig:relative_performance} quantifies the relative improvement of each algorithm over egg. On one hand, we remark the RL methods (Aurora, Omelette) are competitive at guiding the application of rules in equality saturation, with a relative improvement of $99\%$ over egg for $33\%$ of the queries. On the other hand, we remark that Omelette performs worse than egg in two cases. It is important to note that while the relative improvement is reasonably similar between algorithms in some cases (Query 9), the magnitude of improvement differs (Figure \ref{fig:relative_performance}).

Figure \ref{fig:actual_performance} shows the execution latency (left) and planning latency (right) when using Aurora and egg to rewrite TPC-H queries on Risinglight. We observe that a non-trivial application of rules within equality saturation, such as the one learnt by Aurora, greatly improves upon the state-of-the-art equality saturation solver, egg. We set a timeout of 350 seconds for the execution to finish given the rather small size of the data pool (4.7GB). In 4 out of 6 cases, egg runs into a timeout. This is consistent across the 5 trials we test over. The importance of this result is two-fold. First, it is clear that Aurora can find more competitive plans when compared to egg. Second, the cost model we define is generally a good proxy for the true latency. Aurora's planning latency is higher for Query 5 and Query 9. This is expected, as a wise e-graph construction routine should allow for more rewrites while keeping the e-graph within the node limit. However, we emphasize that for this experiment planning operates on an insignificant scale (milliseconds), which combined with our previous result on the execution latency paints a compelling picture to prove that Aurora outputs query plans with better end-to-end query latency.

\subsubsection{Case Study: PostgreSQL}

This section presents a study case on PostgreSQL. The motivation is to show Aurora is DBMS-agnostic. We use Aurora to rewrite the \textit{Alternative} query from Table \ref{tab:pgsql_case} into the \textit{Optimal} form. Subsequently, the rewritten query is executed in PostgreSQL.

\begin{table}[ht]
      \caption{Equivalent queries with different levels of performance.} 
    \label{tab:pgsql_case}
  \begin{tabular}{lp{0.6\linewidth}} 
    \toprule
    Alias & Query\\
    \midrule
    
    Original &  \textit{select o\_orderkey from orders where o\_orderstatus = 'F' and o\_orderkey in (select o\_orderkey from orders where o\_totalprice $>$ 40000)}\\
    
    Optimal & \textit{select o\_orderkey from orders where o\_orderstatus = 'F' and o\_totalprice $>$ 40000}\\
    
    Alternative & \textit{select key from (select o\_orderkey as key, o\_orderstatus as status, o\_totalprice as price from orders where o\_totalprice $>$ 40000 ) as inter where status = 'F' and price $>$ 40000} \\
    \bottomrule
  \end{tabular}
\end{table}

Figure \ref{fig:postgres} shows that Aurora can be generalised across disparate DBMSs. We remark the plans devised by PostgreSQL incur higher latency than the query plan returned by Aurora. In the experiment, Aurora receives the second query, coined \textit{Alternative}, and returns the \textit{Optimal}. The rewritten query is $11\%$ faster. We observe the PostgreSQL optimiser does not take advantage of the correlation between the \texttt{FILTER} operation in the outer query and the one in the nested query when executing the former (i.e., \textit{``price $>$ 40000"}), resulting in higher latency. The \textit{Original} query is added for reference to show Aurora can theoretically bring large improvements over mainstream DBs. However, Aurora currently does not support \textbf{IN} operators over nested queries, while PostgreSQL does. This engineering extension is left for future work.

\subsection{The Importance of RL in Aurora} \label{sec:exp_mot}

\begin{figure} [t]
\includegraphics[scale=0.4]{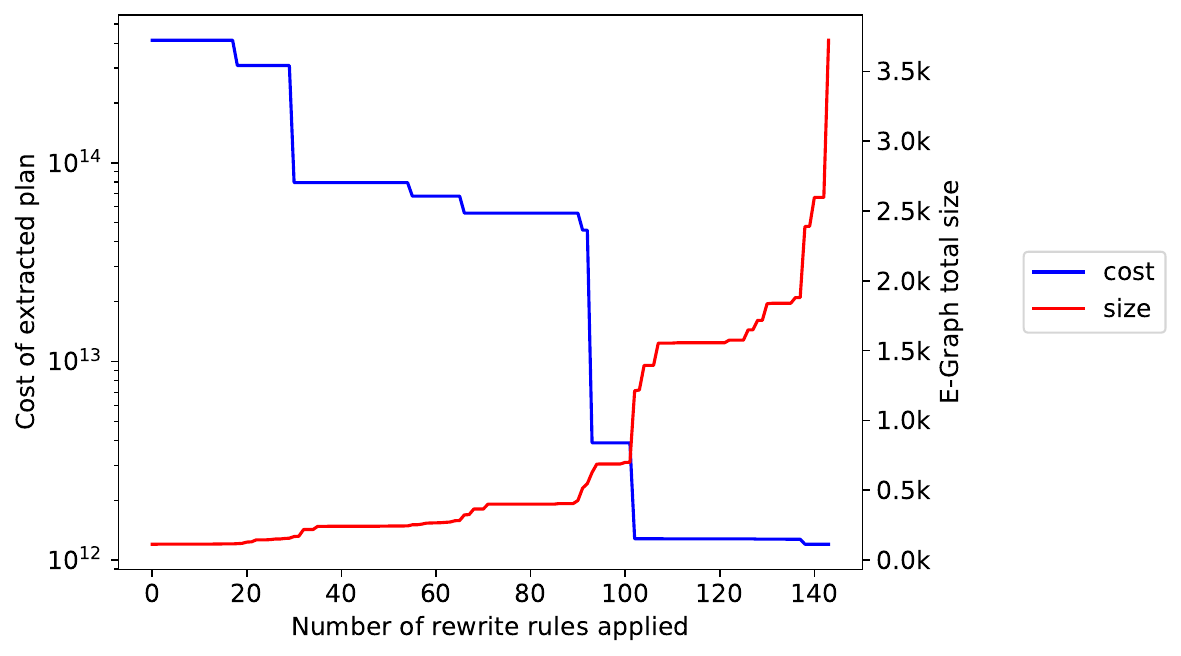}

\centering
\caption{Results for TPC-H Query 5 rewritten using egg. The relation between the size of the e-graph and the estimated cost of the extracted SQL plan. The number of nodes is limited to 3.5k. \textbf{Red line:} the growth trend for the number of nodes in the e-graph. \textbf{Blue line:} the cost of the extracted plan after each rewrite rule.}
\label{fig:cost_to_node}
\end{figure}

This section empirically motivates introducing RL in equality saturation to avoid pressing challenges such as high memory and algorithmic latency.

Figure \ref{fig:cost_to_node} shows the link between the size of the e-graph and the cost of the extracted plan using egg on TPC-H Query 5, a 6-table join query. Aligned with our expectations, as more equivalent query plans are added to the e-graph, the cost of the extracted plan decreases, while the number of nodes in the e-graph increases. A clear observation is that the node increase is not proportional to the cost improvement. For instance, we remark how the size of the e-graph explodes after the first $140$ rule applications, while the cost remains rather constant during the same window. 

Since egg applies the rewrite rules sequentially, we raise the question of whether a plan with a lower cost can be found within the same node limit, by changing the order in which the rules are applied as well as their frequency. This is possible if: 1) the e-graph node limit is a concern and therefore bounded (e.g. egg is set to stop upon reaching 10k nodes by default) and 2) the e-graph does not get saturated within the node limit. Upon reaching saturation, the e-graph encodes all possible plans so the extracted plans will always be optimal.

\begin{figure}[t]
\includegraphics[scale=0.42]{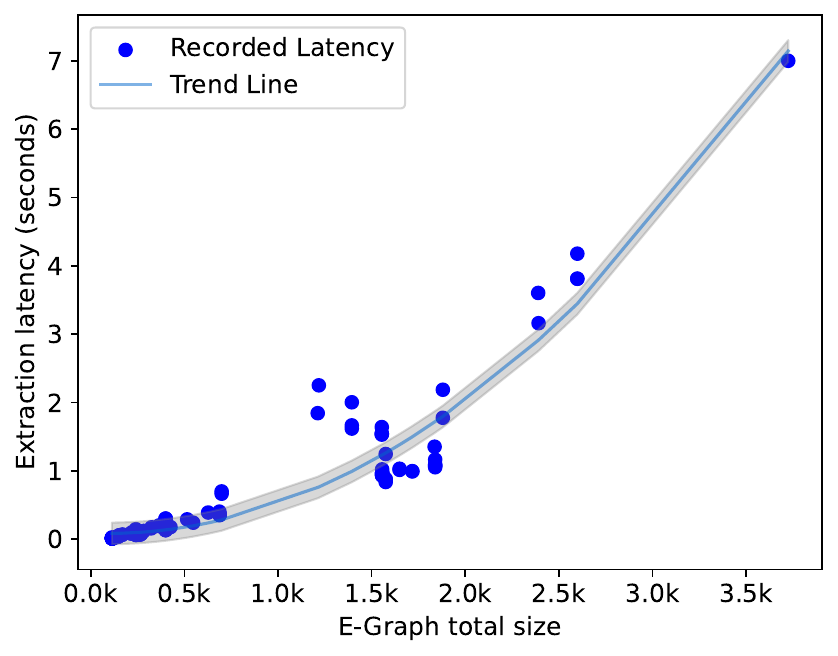} 
\includegraphics[scale=0.42]{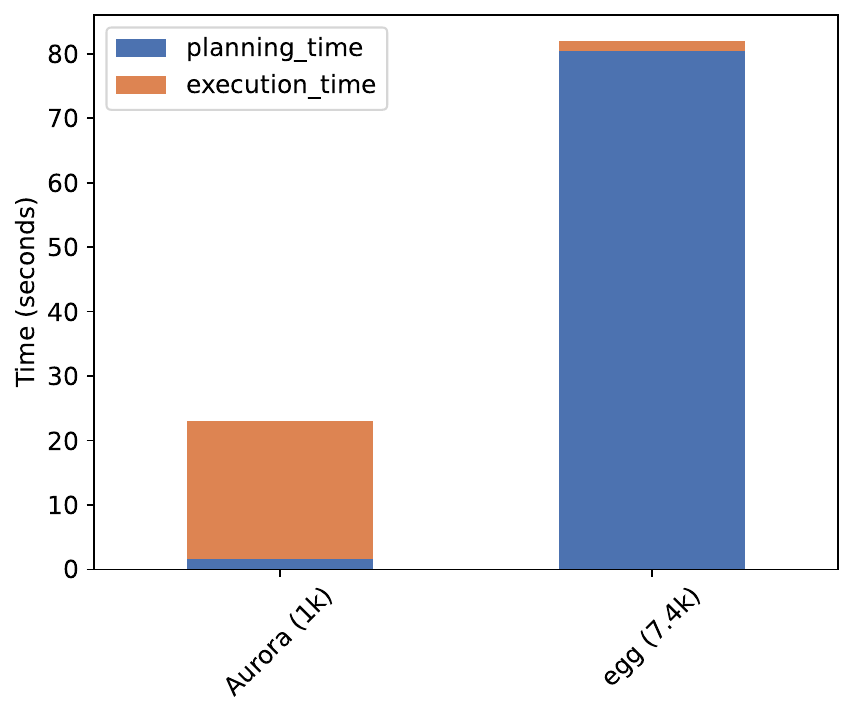} 
\centering
\caption{\textbf{Top:} The relation between the size of the e-graph and the extraction latency. \textbf{Bottom:} Planning and execution statistics for two alternative plans for TPC-H Query 5 executed on Risinglight. The plans are extracted with ILP.} 
\label{fig:limit_2}
\end{figure}

Figure \ref{fig:limit_2} shows the relation between the size of the e-graph and query latency. We observe from the top plot that as the number of equivalent plans grows in the e-graph, extracting the optimal query plan from it incurs a higher cost. In the bottom plot, we measure the planning time (including extraction), as well as the execution time for two Query 5 plans devised with flavours of equality saturation. The execution time is approximately $20$ seconds with a standard deviation of $2s$, and a planning latency in the order of milliseconds using Aurora. The second plan is extracted from an e-graph of $7.3k$ nodes built using egg. In this case, the planning latency has a mean of $78$ seconds, with a large percentage associated with extracting the optimal query plan from the e-graph. The execution time is discarded as the planning time is already over $3$ times higher than the total latency for the first plan. This shows equality saturation cannot be trivially applied to optimise queries, and RL is necessary as per our proposed system, Aurora.

\section{Related Work}

Query optimisation has a long history of research in the database community. Recently, machine learning techniques have been introduced into the query optimisation pipeline at different critical points, ranging from query plan cost estimation \cite{sun2019end, marcus2019plan} to 
 the \textit{join ordering} problem \cite{marcus2019neo, yu2020reinforcement, kadkhodaei2011combination}. Much of the existing work neglects other logical rewrite opportunities. While current methods can efficiently estimate costs and propose competitive join orders, the field is scarce in terms of rewriting complex queries. The reason is two-fold: 1) the space of possible rewritten queries can be enormous, and 2) the ordering problem is very difficult to learn. Of particular importance, we identify the work of \cite{zhou2021learned}, who seek to learn over the space of all query rewrite orders with RL. Our work, which adapts equality saturation for query rewrite is orthogonal, as we propose to reduce the search space by learning over the query e-graph space.

Equality saturation has seen a boost in popularity as a consequence of ever-faster hardware and improvements to the algorithm. E-graphs good (egg) \cite{willsey2021egg} revitalises the  methodology with asymptotic latency improvement over the original solver by \cite{tate2009equality}. Omelette \cite{omelette} extends egg by treating the application of rules as a graph combinatorial problem. Omelette shares the same aim as Aurora, to encode as many competitive plans in the e-graph as possible within a memory limit. Omelette demonstrates graph representation learning can be used to learn e-graphs for simple programming languages, such as first-order logic, a core result which we use in Aurora to tackle query rewrite. 

\section{Conclusion}

In this research paper, we have demonstrated \textit{how} to use equality saturation for query rewrite and the advantages it brings over mainstream optimisers. We have shown that equality saturation cannot trivially optimise queries of moderate complexity due to high planning latency and memory considerations. To address these challenges, we have introduced Aurora, a query optimiser that guides the application of rewrite rules within the equality saturation framework. By directing our attention to learning methods, we have devised an expressive RL pipeline for solving the e-graph construction combinatorial problem, which extends to other programming languages with minimal change.  

\subsection{Limitations \& Future Work}
The most pressing problem in Aurora is generalising across disparate queries. Training an agent to solve NP-hard problems across sparse problem instances (such as the space of possible query e-graphs) is not realistic. We leave as future work the adoption of more advanced RL techniques such as population-based RL \cite{jaderberg2019human} to improve generalisation. 

\bibliography{mlsys2024style/aurora_paper}

\begin{thebibliography}{36}
\providecommand{\natexlab}[1]{#1}
\providecommand{\url}[1]{\texttt{#1}}
\expandafter\ifx\csname urlstyle\endcsname\relax
  \providecommand{\doi}[1]{doi: #1}\else
  \providecommand{\doi}{doi: \begingroup \urlstyle{rm}\Url}\fi

\bibitem[Begoli et~al.(2018)Begoli, Camacho-Rodr{\'\i}guez, Hyde, Mior, and Lemire]{begoli2018apache}
Begoli, E., Camacho-Rodr{\'\i}guez, J., Hyde, J., Mior, M.~J., and Lemire, D.
\newblock Apache calcite: A foundational framework for optimized query processing over heterogeneous data sources.
\newblock In \emph{Proceedings of the 2018 International Conference on Management of Data}, pp.\  221--230, 2018.

\bibitem[Finance \& Gardarin(1991)Finance and Gardarin]{finance1991rule}
Finance, B. and Gardarin, G.
\newblock A rule-based query rewriter in an extensible dbms.
\newblock In \emph{Proceedings. Seventh International Conference on Data Engineering}, pp.\  248--249. IEEE Computer Society, 1991.

\bibitem[Frydenberg(1990)]{frydenberg1990chain}
Frydenberg, M.
\newblock The chain graph markov property.
\newblock \emph{Scandinavian Journal of Statistics}, pp.\  333--353, 1990.

\bibitem[Hochreiter(1998)]{hochreiter1998vanishing}
Hochreiter, S.
\newblock The vanishing gradient problem during learning recurrent neural nets and problem solutions.
\newblock \emph{International Journal of Uncertainty, Fuzziness and Knowledge-Based Systems}, 6\penalty0 (02):\penalty0 107--116, 1998.

\bibitem[Hochreiter \& Schmidhuber(1997)Hochreiter and Schmidhuber]{hochreiter1997long}
Hochreiter, S. and Schmidhuber, J.
\newblock Long short-term memory.
\newblock \emph{Neural computation}, 9\penalty0 (8):\penalty0 1735--1780, 1997.

\bibitem[Huang et~al.(2022)Huang, Dossa, Raffin, Kanervisto, and Wang]{huang202237}
Huang, S., Dossa, R. F.~J., Raffin, A., Kanervisto, A., and Wang, W.
\newblock The 37 implementation details of proximal policy optimization.
\newblock \emph{The ICLR Blog Track 2023}, 2022.

\bibitem[Jaderberg et~al.(2019)Jaderberg, Czarnecki, Dunning, Marris, Lever, Castaneda, Beattie, Rabinowitz, Morcos, Ruderman, et~al.]{jaderberg2019human}
Jaderberg, M., Czarnecki, W.~M., Dunning, I., Marris, L., Lever, G., Castaneda, A.~G., Beattie, C., Rabinowitz, N.~C., Morcos, A.~S., Ruderman, A., et~al.
\newblock Human-level performance in 3d multiplayer games with population-based reinforcement learning.
\newblock \emph{Science}, 364\penalty0 (6443):\penalty0 859--865, 2019.

\bibitem[Kadkhodaei \& Mahmoudi(2011)Kadkhodaei and Mahmoudi]{kadkhodaei2011combination}
Kadkhodaei, H. and Mahmoudi, F.
\newblock A combination method for join ordering problem in relational databases using genetic algorithm and ant colony.
\newblock In \emph{2011 IEEE International Conference on Granular Computing}, pp.\  312--317. IEEE, 2011.

\bibitem[Kipf \& Welling(2016)Kipf and Welling]{kipf2016semi}
Kipf, T.~N. and Welling, M.
\newblock Semi-supervised classification with graph convolutional networks.
\newblock \emph{arXiv preprint arXiv:1609.02907}, 2016.

\bibitem[Marcus \& Papaemmanouil(2019)Marcus and Papaemmanouil]{marcus2019plan}
Marcus, R. and Papaemmanouil, O.
\newblock Plan-structured deep neural network models for query performance prediction.
\newblock \emph{arXiv preprint arXiv:1902.00132}, 2019.

\bibitem[Marcus et~al.(2019)Marcus, Negi, Mao, Zhang, Alizadeh, Kraska, Papaemmanouil, and Tatbul]{marcus2019neo}
Marcus, R., Negi, P., Mao, H., Zhang, C., Alizadeh, M., Kraska, T., Papaemmanouil, O., and Tatbul, N.
\newblock Neo: A learned query optimizer.
\newblock \emph{arXiv preprint arXiv:1904.03711}, 2019.

\bibitem[Medsker \& Jain(2001)Medsker and Jain]{medsker2001recurrent}
Medsker, L.~R. and Jain, L.
\newblock Recurrent neural networks.
\newblock \emph{Design and Applications}, 5:\penalty0 64--67, 2001.

\bibitem[Nelson \& Oppen(1980)Nelson and Oppen]{nelson1980fast}
Nelson, G. and Oppen, D.~C.
\newblock Fast decision procedures based on congruence closure.
\newblock \emph{Journal of the ACM (JACM)}, 27\penalty0 (2):\penalty0 356--364, 1980.

\bibitem[Nieuwenhuis \& Oliveras(2005)Nieuwenhuis and Oliveras]{nieuwenhuis2005proof}
Nieuwenhuis, R. and Oliveras, A.
\newblock Proof-producing congruence closure.
\newblock In \emph{Term Rewriting and Applications: 16th International Conference, RTA 2005, Nara, Japan, April 19-21, 2005. Proceedings 16}, pp.\  453--468. Springer, 2005.

\bibitem[Norris(1998)]{norris1998markov}
Norris, J.~R.
\newblock \emph{Markov chains}.
\newblock Number~2. Cambridge university press, 1998.

\bibitem[Piao \& Liu(2019)Piao and Liu]{piao2019energy}
Piao, C. and Liu, C.~H.
\newblock Energy-efficient mobile crowdsensing by unmanned vehicles: A sequential deep reinforcement learning approach.
\newblock \emph{IEEE Internet of Things Journal}, 7\penalty0 (7):\penalty0 6312--6324, 2019.

\bibitem[Pirahesh et~al.(1992)Pirahesh, Hellerstein, and Hasan]{pirahesh1992extensible}
Pirahesh, H., Hellerstein, J.~M., and Hasan, W.
\newblock Extensible/rule based query rewrite optimization in starburst.
\newblock \emph{ACM Sigmod Record}, 21\penalty0 (2):\penalty0 39--48, 1992.

\bibitem[Poess \& Floyd(2000)Poess and Floyd]{poess2000new}
Poess, M. and Floyd, C.
\newblock New tpc benchmarks for decision support and web commerce.
\newblock \emph{ACM Sigmod Record}, 29\penalty0 (4):\penalty0 64--71, 2000.

\bibitem[PostgreSQL(1996)]{postgresql1996postgresql}
PostgreSQL, B.
\newblock Postgresql.
\newblock \emph{Web resource: http://www. PostgreSQL. org/about}, 1996.

\bibitem[Puterman(2014)]{puterman2014markov}
Puterman, M.~L.
\newblock \emph{Markov decision processes: discrete stochastic dynamic programming}.
\newblock John Wiley \& Sons, 2014.

\bibitem[Risinglight(2022)]{rising}
Risinglight.
\newblock Risinglight.
\newblock \url{https://github.com/risinglightdb/risinglight}, 2022.
\newblock Accessed: 10 01, 2023.

\bibitem[Schulman et~al.(2017)Schulman, Wolski, Dhariwal, Radford, and Klimov]{schulman2017proximal}
Schulman, J., Wolski, F., Dhariwal, P., Radford, A., and Klimov, O.
\newblock Proximal policy optimization algorithms.
\newblock \emph{arXiv preprint arXiv:1707.06347}, 2017.

\bibitem[Sherstinsky(2020)]{sherstinsky2020fundamentals}
Sherstinsky, A.
\newblock Fundamentals of recurrent neural network (rnn) and long short-term memory (lstm) network.
\newblock \emph{Physica D: Nonlinear Phenomena}, 404:\penalty0 132306, 2020.

\bibitem[Singh(2022)]{omelette}
Singh, Z.
\newblock Deep reinforcement learning for equality saturation.
\newblock \url{https://www.cl.cam.ac.uk/~ey204/pubs/MPHIL_P3/2022_Zak.pdf}, 2022.
\newblock Accessed: 06 09, 2022.

\bibitem[Sun \& Li(2019)Sun and Li]{sun2019end}
Sun, J. and Li, G.
\newblock An end-to-end learning-based cost estimator.
\newblock \emph{arXiv preprint arXiv:1906.02560}, 2019.

\bibitem[Sutton \& Barto(2018)Sutton and Barto]{sutton2018reinforcement}
Sutton, R.~S. and Barto, A.~G.
\newblock \emph{Reinforcement learning: An introduction}.
\newblock MIT press, 2018.

\bibitem[Tarjan(1975)]{tarjan1975efficiency}
Tarjan, R.~E.
\newblock Efficiency of a good but not linear set union algorithm.
\newblock \emph{Journal of the ACM (JACM)}, 22\penalty0 (2):\penalty0 215--225, 1975.

\bibitem[Tate et~al.(2009)Tate, Stepp, Tatlock, and Lerner]{tate2009equality}
Tate, R., Stepp, M., Tatlock, Z., and Lerner, S.
\newblock Equality saturation: a new approach to optimization.
\newblock In \emph{Proceedings of the 36th annual ACM SIGPLAN-SIGACT symposium on Principles of programming languages}, pp.\  264--276, 2009.

\bibitem[Velickovic et~al.(2017)Velickovic, Cucurull, Casanova, Romero, Lio, Bengio, et~al.]{velickovic2017graph}
Velickovic, P., Cucurull, G., Casanova, A., Romero, A., Lio, P., Bengio, Y., et~al.
\newblock Graph attention networks.
\newblock \emph{stat}, 1050\penalty0 (20):\penalty0 10--48550, 2017.

\bibitem[Wang et~al.(2020)Wang, Hutchison, Leang, Howe, and Suciu]{wang2020spores}
Wang, Y.~R., Hutchison, S., Leang, J., Howe, B., and Suciu, D.
\newblock Spores: sum-product optimization via relational equality saturation for large scale linear algebra.
\newblock \emph{arXiv preprint arXiv:2002.07951}, 2020.

\bibitem[Wang et~al.(2022)Wang, Khamis, Ngo, Pichler, and Suciu]{wang2022optimizing}
Wang, Y.~R., Khamis, M.~A., Ngo, H.~Q., Pichler, R., and Suciu, D.
\newblock Optimizing recursive queries with program synthesis.
\newblock \emph{arXiv preprint arXiv:2202.10390}, 2022.

\bibitem[Willsey et~al.(2021)Willsey, Nandi, Wang, Flatt, Tatlock, and Panchekha]{willsey2021egg}
Willsey, M., Nandi, C., Wang, Y.~R., Flatt, O., Tatlock, Z., and Panchekha, P.
\newblock Egg: Fast and extensible equality saturation.
\newblock \emph{Proceedings of the ACM on Programming Languages}, 5\penalty0 (POPL):\penalty0 1--29, 2021.

\bibitem[Yang et~al.(2021)Yang, Phothilimthana, Wang, Willsey, Roy, and Pienaar]{yang2021equality}
Yang, Y., Phothilimthana, P., Wang, Y., Willsey, M., Roy, S., and Pienaar, J.
\newblock Equality saturation for tensor graph superoptimization.
\newblock \emph{Proceedings of Machine Learning and Systems}, 3:\penalty0 255--268, 2021.

\bibitem[Yu et~al.(2020)Yu, Li, Chai, and Tang]{yu2020reinforcement}
Yu, X., Li, G., Chai, C., and Tang, N.
\newblock Reinforcement learning with tree-lstm for join order selection.
\newblock In \emph{2020 IEEE 36th International Conference on Data Engineering (ICDE)}, pp.\  1297--1308. IEEE, 2020.

\bibitem[Zhou et~al.(2020)Zhou, Cui, Hu, Zhang, Yang, Liu, Wang, Li, and Sun]{zhou2020graph}
Zhou, J., Cui, G., Hu, S., Zhang, Z., Yang, C., Liu, Z., Wang, L., Li, C., and Sun, M.
\newblock Graph neural networks: A review of methods and applications.
\newblock \emph{AI open}, 1:\penalty0 57--81, 2020.

\bibitem[Zhou et~al.(2021)Zhou, Li, Chai, and Feng]{zhou2021learned}
Zhou, X., Li, G., Chai, C., and Feng, J.
\newblock A learned query rewrite system using monte carlo tree search.
\newblock \emph{Proceedings of the VLDB Endowment}, 15\penalty0 (1):\penalty0 46--58, 2021.

\end{thebibliography}
\bibliographystyle{mlsys2024}

\end{document}